# Disorder dependent spin-orbit torques in $L1_0$ FePt single layer

S. Q. Zheng, K. K. Meng[*], Q. B. Liu, J. K. Chen, J. Miao, X. G. Xu, Y. Jiang[**]

*Beijing Advanced Innovation Center for Materials Genome Engineering, School of Materials Science and Engineering, University of Science and Technology Beijing, Beijing 100083, China*

**Abstract:**

We report spin-orbit torques (SOT) in $L1_0$-ordered perpendicularly magnetized FePt single layer, which is significantly influenced by disorder. Recently, self-induced SOT in $L1_0$-FePt single layer has been investigated, which is ascribed to the composition gradient along the film normal direction. However, the determined mechanisms for magnetization switching have not been fully studied. With varying growth temperatures, we have prepared FePt single layers with same thickness (3 nm) but with different disordering. We have found that nearly full magnetization switching only happens in more disordered films, and the magnetization switching ratio becomes smaller as increasing $L1_0$ ordering. The method for deriving effective spin torque fields in the previous studies cannot fully explain the spin current generation and self-induced SOT in $L1_0$-FePt single layer. Combined with Magneto-Optical Kerr Effect microscopy and anomalous Hall effect measurements, we concluded that the disorder should determine the formation of domain walls, as well as the spin current generation.

Authors to whom correspondence should be addressed: *kkmeng@ustb.edu.cn, **yjiang@ustb.edu.cn

Current-induced spin-orbit torques (SOT), which mediates the transfer of angular momentum from the lattice to the spin system in magnetic materials and heterostructures with strong spin-orbit coupling (SOC) have been widely investigated recently. The SOT can provide efficient and versatile ways to control magnetization dynamics in different classes of materials, including ferromagnets (FM), antiferromagnets, ferrimagnets, magnetic insulators and FM/heavy metal (HM) heterostructures.[1-8] Two main model mechanisms have been proposed to generate SOT, including the spin Hall effect (SHE) and the Rashba-Edelstein effect.[9-15] In general, the SOT has been most extensively studied experimentally in the HM/FM bilayer structures with interface-induced perpendicular magnetic anisotropy (PMA).[16,17] The charge current flowing through HM film can be converted into spin current due to SHE, which can diffuse into the adjacent FM film and exert torques on the magnetization of the FM layer. By this way, the magnitude of SOT induced by SHE is determined by the spin Hall angle of the HM film. On the other hand, the interfacial Rashba effect produced by the structural inversion asymmetry can also give rise to SOT, for which the accumulated spins at the interface can force the moments to change their direction by direct exchange coupling. However, both SHE and Rashba effect overlook the details of the interfacial orbital overlap, which can be quite subtle in HM/FM interfaces and lead to enhanced orbital magnetization and related spin-orbit phenomena.[15] Moreover, in the commonly used HM/FM bilayer structures, the distinction of spin current generated by "bulk"-SHE and "interface"-Rashba effect remains principally blurred. To overcome these issues, SOT has also been found and investigated in single FM layers such as (Ga,Mn)As[18] and (Ge,Mn)Te[19] with bulk inversion asymmetry, and in single antiferromagnets such as CuMnAs[20] and $Mn_2Au$[21] with locally broken inversion symmetry.

For a FM single layer, it is well known that spin current can be induced by an anomalous Hall effect (AHE), whose polarization should be always parallel to magnetization.[22–24] Generally, the accumulated spins produced through AHE are unable to exert torques on its own magnetization. However, if FM had strong SOC and inversion symmetry breaking, the transverse spin current generated by this SHE is regarded to be magnetization independent and its polarization is perpendicular to both current and its flowing direction.[25-29] Thus, it is proposed that the spin current induced by SHE could also exert torques and achieve magnetization switching on itself, and recent studies have implied the possibility of generating SOT in single ferrimagnetic alloys or FMs.[27, 30, 31] $L1_0$-ordered FePt is one of this typical example but with centrosymmetric crystalline structures, in which the strong SOC in combination with uniaxial symmetry of the tetragonal crystal structure leads to a gigantic bulk PMA.[32-34] The SOT of $L1_0$-FePt single layer has been found recently, which is ascribed to a new type of inversion asymmetry, the composition gradient.[35, 36] The SOT efficiency in $L1_0$-FePt films is found to be increased with enhancing inversion symmetry breaking. However, the microscopic mechanisms that determine the spin current generation and SOT efficiency have not been fully studied. In this paper, we have prepared 3-nm-thick FePt single layers with varying growth temperatures, resulting in a different disordering. As discussed in the previous works,[35, 36], all the films should have same and small composition gradient. However, we have found that nearly full magnetization switching only happens in more disordered films, and the magnetization switching ratio becomes smaller as increasing $L1_0$ ordering. The method for deriving effective spin torque fields in the previous studies seems not to be applicable for investigating self-induced SOT in $L1_0$ FePt single layer. Combined with Magneto-Optical Kerr Effect (MOKE) and AHE measurements, we concluded that the disorder should determine the formation of domain walls, as well as the spin current generation during the domain wall motions.

The 3-nm-thick FePt films were epitaxially deposited on MgO (001) substrates by magnetron co-sputtering from Fe and Pt elemental targets, and the base pressure of the sputtering chamber was less than $7 \times 10^{-6}$ Pa. The substrates were kept at different growth temperatures varying from 350 $^o$C to 450 $^o$C. After deposition, the 3-nm-thick

FePt films were annealed in situ at the same temperature for 2 hours, and then the films were left to cool to room temperature in situ. Figure 1(a) shows a typical cross-sectional transmission electron microscopy (TEM) image of FePt film grown on MgO substrate at 450 $^o$C. There is no magnetic dead layer at interface and the FePt film shows high crystallinity. Figure 1(b) shows the structure characterization of 20-nm-thick FePt films grown at 350 $^o$C, 380 $^o$C, and 450 $^o$C through X-ray diffraction (XRD) measurements. Besides the MgO (002) peaks, the face-centered-tetragonal (fct) (001), (002) and (003) peaks of FePt films in the spectrum become more evident as increasing growth temperature, indicating an enhanced $L1_0$ ordering. The chemical ordering $S$ was calculated from the intensities of (001) and (002) peaks, and the value for FePt films grown at 380 $^o$C and 450 $^o$C are 0.54 and 0.6, respectively. The XRD results for 3-nm-thick FePt films grown at 380 $^o$C and 450 $^o$C have been shown in Fig. S1 of supplementary materials. The thickness and the surface roughness of 3-nm-thick FePt films were been characterized by x-ray reflectivity (XRR) and atomic force microscope (AFM). According to the XRR results of the film grown at 450 $^o$C as shown in Fig. 1(c), the thickness of FePt is 3.13 nm. The roughness is less than 0.32 nm, in agreement with the AFM result as shown in Fig. 1(d). Fig. 2 shows the magnetic hysteresis loops and the Magneto-Optical Kerr Effect (MOKE) results for FePt films grown at 380 $^o$C and 450 $^o$C. For MOKE measurements, the magnetic field is applied perpendicular to the film plane, and a linearly polarized laser beam at normal incidence is focused on the film and reflected with a small rotation angle in polarization. Figs. 2(a) and (b) show the in-plane and out-of-plane magnetic hysteresis loops of the two films measured by VSM, which all show evident bulk PMA with different coercivity. The Kerr images for the two films denoted as 1, 2, 3 (380 $^o$C) and 1', 2', 3' (450 $^o$C) marked in hysteresis loops are shown in Fig. 2(c)-2(e) and Fig. 2(f)-2(h), respectively. It is found that, at remanent state, the film grown at 380 $^o$C shows obscure worm-like configurations in the so-called labyrinthine state, while several bubble-like magnetic domains have been found in the film grown at 450 $^o$C. Generally, the dipolar energy can be saved by breaking the films into domains, but each domain created costs energy because of the cost of the domain walls. Therefore, the size of a domain wall is a balance between the exchange and anisotropy energies, so the formation of domains is a balance between the cost of a demagnetizing field and the cost of a domain wall. The domain wall size is therefore given by $\Delta = \pi\sqrt{A/K}$ and the energy per unit area is $\sigma = \pi\sqrt{AK}$, where $A$ is the exchange constant and $K$ is the effective anisotropy energy. The saturation magnetization $M_s$ and anisotropy field $H_k$ of the two films are shown in Table I. Assuming the same $A$ for the two $L1_0$ ordered films, they should reveal similar domain type and size. However, the MOKE results reveal that there should be other factors for the formation of domain walls. All real materials are inhomogeneous, and magnetization reversal is initiated in a small nucleation volume around a defect. These defects often act as nucleation centre because, in the second quadrant of the hysteresis loop as shown in Figs. 2 (a) and (b), the reverse magnetic field is enhanced in their vicinity. Once a small nucleus has formed, the wall may propagate outwards, growing from the nucleation volume. Otherwise the new wall may become pinned at some other defect. Besides the competition between exchange and anisotropy energies, the varied amounts and types of defects with changing growth temperatures should also determine the domain structures in the FePt films. Considering the surface roughness of the film grown at 380 $^o$C is 0.38 nm as shown in Fig. S2 of the supplementary materials, which is similar with 450 $^o$C film, we suppose that the two films have similar surface defects. Therefore, one possible type of defects that determines the domain structures is the short-range ordering, which is a tendency to locate a Pt antisite defect next to an Fe antisite defect.[37] Although the chemical ordering $S$ is larger for FePt films grown at 450 $^o$C, this kind of defects become more randomly distributed and play a more important role for domain walls nucleation and expansion.

Then, the current induced magnetization switching measurements have been carried out. The Hall bar of 20 μm×120 μm were patterned for electrical measurement by electron beam lithography and Ar ion milling, and a lift-off process were used to form contact electrodes. The schematic of the Hall bar along with the definition of the coordinate system used in this study is shown in Fig. 3(a). Figure 3(b)-3(e) show the magnetic field dependence of anomalous Hall resistance (*R*-***H***) and current-induced magnetization switching behaviors (*R*-***J***) at room temperature for $L1_0$ FePt film with varying growth temperatures ($T_g$) from 350 °C to 450 °C. The current-induced magnetization switching behaviors in $L1_0$ FePt films were measured by applying a pulsed current with the width of 50 μs, and the resistance was measured after a 16 μs delay under an in-plane field along applied current directions at room temperature. As shown in supplementary Figs. S3(a) and (b), reversible current-induced magnetization switching of $L1_0$ FePt film with $T_g$=380 °C are achieved with ***H*** = ±500 Oe and the critical switching current $I_c$ is 30 mA. Opposite switching polarities are observed when ***H*** is reversed, indicating a typical current-induced magnetization switching behavior similar with the SOT switching reported in previous which suggesting a similar underlying driving torque exist in $L1_0$ FePt film. However, as compared with the change of Hall resistances, we have found that fewer part of magnetization can be switched in $L1_0$ FePt films as increasing growth temperatures. The relationship between magnetization switching ratio and growth temperature has been summarized in Fig. 3(f). It is quite similar to the partial switching observed in the previous SOT research of $L1_0$ FePt films.[35, 36] MOKE microscopy were also carried out to characterize the magnetization switching driven by pulsed current for the films grown at 380 °C and 450 °C as shown in Fig. S4 of supplementary materials. Compared with the SOT in HM/FM bilayers, this partial magnetization switching is far more difficult and not well explored.

To quantitatively investigate the spin-torque efficiency in $L1_0$ FePt films, the harmonic measurements with sweeping small in-plane magnetic fields ***$H_x$*** and ***$H_y$*** (parallel or perpendicular to the current direction) were carried out at room temperature. The first $V_\omega$ and second $V_{2\omega}$ harmonic Hall voltages signals of $L1_0$ FePt films were detected by two lock-in amplifier systems simultaneously. Before the harmonic measurements, we have applied a large out-of-plane external field $M_z$ to saturate the magnetization of two films, which remain saturated after the field is turned off. The selected results of the films with $T_g$=380 °C and $T_g$=450 °C measurements by applying a sinusoidal AC current with an amplitude of 3 mA were plotted against small magnetic fields in Fig. 4, in which the signals are measured with the out-of-plane magnetization component $M_z > 0$ and $M_z < 0$. The results of the film grown at 380 °C are shown in Fig. S5 of supplementary materials. The second harmonic Hall voltages against $H_y$ for $L1_0$ FePt film with $T_g$=450 °C are too weak to be captured. Even for $L1_0$ FePt film with $T_g$=380 °C, the second harmonic Hall voltages against $H_y$ are also not strong enough to exhibit excellent linearity. The damping-like effective fields $H_D$ and field-like effective fields $H_F$ can be calculated by the following equation:[38-40]

$$H_{D(F)} = -2\frac{H_{L(T)} \pm 2\xi H_{T(L)}}{1-4\xi^2} \quad (1)$$

where the $\xi$ is the ratio of planar Hall effect (PHE) resistance and anomalous Hall effect (AHE) resistance and the $\pm$ sign is the magnetization pointing $\pm Z$. The $H_{L(T)}$ can be determined using the following equation:

$$H_{L(T)} = \frac{\partial V_{2\omega}/\partial H_{x(y)}}{\partial^2 V_\omega/\partial H_{x(y)}^2} \quad (2)$$

The value of $\xi$ have been determined through the PHE measurement under an in-plane magnetic field of 9 T.[40, 41] The value of $\xi$ for $L1_0$ FePt films with $T_g$=380 °C and $T_g$=450 °C are 0.058 and 0.13. Using equations (1) and (2), $H_D$ and $H_F$ for FePt films with varying growth temperatures from 350 °C to 450 °C are shown in Fig. 5.

Here, we will focus on the results of the films with $T_g$=380 °C and $T_g$=450 °C. Liu et al.[35] have reported the spin torque effective field could monotonously increase to six times with the value of $S$ increasing from 0.6 to 0.75. In addition, Tang et al.[36] have reported the spin torque effective field could monotonously increase to eight times with the value of $S$ increasing from 0.57 to 0.91. For $L1_0$ FePt films with with $T_g$=380 °C and $T_g$=450 °C, the value of $H_D$ increases sharply from 5.41 to 24.42 Oe which nearly increased five times as $S$ increased 6%. However, it should be noted that, although largest $H_D$ have been found for this film, the magnetization switching ratio is smallest for $T_g$=450 °C as shown in Fig. 3(f). This great variation cannot be explained only by the change of $S$ and composition gradient, and the influence of other factors on the self-induced SOT $L1_0$ FePt single layer should be further investigated. On the other hand, thermal effects including anomalous Nernst effect and spin Seebeck effect have no dramatic influence on the result of effective spin-torque fields and can be neglected as shown in Figs. S6 and S7 of supplementary materials. Since all the films are grown on the same substrate and have the same thickness, the different influence coming from Rashba effect can be ruled out. Actually, the derived $H_D$ and $H_F$ are too small to adequately account for the magnetization switching of $L1_0$ FePt. Tang et al. ascribed this discrepancy to the fact that the effective fields are derived for magnetically saturated states, while the current induced magnetization switching is achieved through domain nucleation and expansion processes. On the other hand, the thicknesses of FePt films in our work are all 3 nm, which should have small composition gradient.[36] Furthermore, as shown in Fig. 3, the largest magnetization switching ratio happens in most disordered films. According to the MOKE measurements, the defects such as short-range ordering should determine the domain structures and domain wall motions. On the other hand, the thicknesses of FePt films in our work are all 3 nm, which should have small composition gradient.[36] Furthermore, as shown in Fig. 3, the largest magnetization switching ratio happens in most disordered films.

To further explore the disorder dependent spin transport in FePt films, we have also investigated the AHE. Here, we also focus on the films with $T_g$=380 $^{\circ}$C and $T_g$=450 $^{\circ}$C. Figs. 6(a) and (b) show the measured $\rho_{xx}$ and $\rho_{xy}$ for two films by PPMS as a function of temperature ($T$), which all decreases as cooling down. The temperature dependence of $\rho_{xx}$ can be well fitted by $T^2$ across a wide temperature range as shown in Fig. 6(a). It reveals that spin flip is the dominant scattering mechanism, because the one-magnon scattering process leads to the $T^2$ dependence of the resistivity when there are spin-up and spin-down electrons at the Fermi surface.[42–44] We plot $\rho_{xy}$ as a function of $\rho^2_{xx}$ for two films to compare the difference of scaling curves in Figs. 6(c) and (d). Conventionally, three relatively distinct contributions to the Hall effect are discussed: the so-called intrinsic Berry phase contribution stemming from the electronic structure of a pristine crystal, and two extrinsic contributions which arise from disorder, namely, the side jump and skew scattering.[45, 46] Recent investigations have explored the possible mechanisms of AHE through giving different scaling laws describing $\rho_{xy}$ in terms of $\rho_{xx}$.[47, 48] Here, we have used two scaling laws to explore the contributions of AHE. Firstly, it is found that $\rho_{xy}$ can be described by scaling law $\rho_{xy}=a\rho_{xx0}+b\rho_{xx}^2$ as shown in Fig. 6(a) and 6(b), where $\rho_{xx0}$ is the residual resistivity induced by impurity scattering, $a$ is the skew scattering coefficient, and the coefficient $b$ contain intrinsic Berry curvature and side-jump contributions.[47] Table I shows the value of $a$ and $b$ of $L1_0$ FePt films with $T_g$=380 $^{\circ}$C and $T_g$=450 $^{\circ}$C. It is found that $a$ change its sign with $T_g$ from 380 $^{\circ}$C to 450 $^{\circ}$C, which should be due to the extrinsic contribution from skew scattering. A similar sign change of $a$ is also observed in epitaxial Fe films with film thickness,[47] Co/Pd multilayers with bilayer repetition[49] and previous studies of $L1_0$ FePt films.[34, 50] On the contrary, the magnitude of $b$ changes less with $T_g$, which we consider that both the intrinsic Berry curvature and side-jump contributions are almost unchanged as increasing $T_g$. This result is consistent with the evaluation by first-principles, which reported that the intrinsic and side-jump contributions are almost constant with different $S$.[51] Skew scattering is due to chiral features which appear in the disorder scattering of spin-orbit coupled ferromagnets.[47] Therefore, even though $L1_0$ FePt films with $T_g$=380 °C and $T_g$=450 °C have nearly the same $L1_0$ ordering, the skew scattering can be dramatically varied due to the distribution, the type, and the density of disorder. The AHE analysis based on the other scaling law has been shown in Fig. S8 and Table SI of supplementary materials.[48] The results indicate that the disorder should make a significant contribution to the spin transport of FePt films. The most disordered FePt films have the largest magnetization switching ratio but the derived spin torque fields are very small. The above method for deriving effective spin torque fields seems not to be applicable, and a larger spin current should be generated in more disordered films, which may be represented by skew scattering since SHE shares similar physics and mechanisms of the AHE. A thorough understanding of the complete physical phenomenon still requires further research.

In summary, we have compared and investigated the current induced magnetization switching in 3-nm-thick FePt single layer with different growth temperatures. An enhanced $L1_0$ ordering has been induced as increasing growth temperatures. MOKE measurements revealed that the domain wall formation should be dramatically determined by the defects such as short-range ordering. We have found that nearly full magnetization switching only happens in more disordered films, and the magnetization switching ratio becomes smaller as increasing $L1_0$ ordering. The method for deriving effective spin torque fields in the previous studies cannot fully explain the spin current generation and self-induced SOT in $L1_0$-FePt single layer. We concluded that the disorder should determine the formation of domain walls, as well as the spin current generation.

**See more detailed discussion in the Supplementary materials.**

**Data Availability Statement:** The data that supports the findings of this study are available within the article [and its supplementary material].

**Acknowledgements:** This work was partially supported by the National Natural Science Foundation of China (Grant Nos. 51971027, 51971024, 51927802, 51971023, 51971027, 51731003), and the Fundamental Research Funds for the Central Universities (FRF-TP-19-001A3, FRF-BD-18-014A).

**Figure captions**

**FIG. 1.** (a) Transmission electron microscopy image of 20 nm FePt film grown at 450 $^{o}$C on MgO substrate. (b) X-ray diffraction of the 20 nm FePt films with varying growth temperatures. (c) X-ray reflectivity of 3 nm FePt films grown at 450 $^{o}$C. (d)

Atomic force microscope of 3 nm FePt films grown at 450 °C.

**FIG. 2.** In-plane and out-of-plane magnetic hysteresis loops of the FePt films grown at 380 °C (a) and 450 °C (b) measured by VSM. Selected Kerr images denoted as 1, 2, 3 (380 °C) and 1', 2', 3' (450 °C) are shown in (c)-(e) and (f)-(h), respectively. The scale bar is 20 μm.

**FIG. 3.** (a) Schematic of the Hall bar along with the definition of the coordinate system used in this study. (b)-(e) Magnetic field dependence of anomalous Hall resistance (*R-H*) and current-induced magnetization switching behaviors (*R-J*) for $L1_0$ FePt films with varying growth temperatures from 350 °C to 450 °C. (f) The relationship between magnetization switching ratio and growth temperature.

**FIG. 4.** (a)-(d) The first $V_\omega$ and second $V_{2\omega}$ harmonic Hall voltages plotted against the small in-plane external fields $H_x$ and $H_y$ in $L1_0$ FePt film with $T_g$=380 °C. (e)-(f) $V_\omega$ and $V_{2\omega}$ plotted against the small in-plane external fields $H_x$ in $L1_0$ FePt film with $T_g$=450 °C. The black and red signals are measured with the out-of-plane magnetization component $M_z > 0$ and $M_z < 0$, respectively.

**FIG. 5.** The relationship between spin torque effective fields and growth temperatures.

**FIG. 6.** Temperature dependence of the longitudinal resistivity $\rho_{xx}$ (a) and transverse resistivity $\rho_{xy}$ (b) for films with with $T_g$=380 °C and with $T_g$=450 °C. The $\rho_{xy}$ as the function of $\rho^2_{xx}$ for $L1_0$ FePt film with $T_g$=380 °C (c) and $L1_0$ FePt film with $T_g$=450 °C (d). The red lines are fitting results with $\rho_{xy}=a\rho_{xx0}+b\rho_{xx}^2$ for two films.

**TABLE I.** Growth temperature $T_g$, saturation magnetizations $M_s$, anisotropy field $H_k$, effective anisotropy energy $K$, magnetization switching ratio, skew scattering coefficient $a$, intrinsic coefficient $b$ for two $L1_0$ FePt films.

| $T_g$ (°C) | $M_s$ (kA/m) | $H_k$ (T) | $K$ (kJ/m$^3$) | Switching ratio | $a$ | $b$ ($\mu\Omega^{-1}$cm$^{-1}$) |
|---|---|---|---|---|---|---|
| 380 | 1240 | 0.8 | 496 | 33% | $2.02 \times 10^{-3}$ | $4.29 \times 10^{-4}$ |
| 450 | 1190 | 1 | 595 | 2.5% | $-2 \times 10^{-3}$ | $2.32 \times 10^{-4}$ |

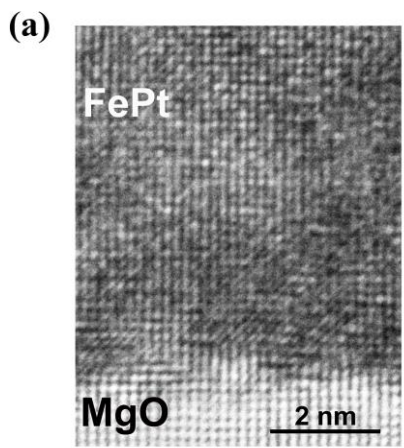

(a)

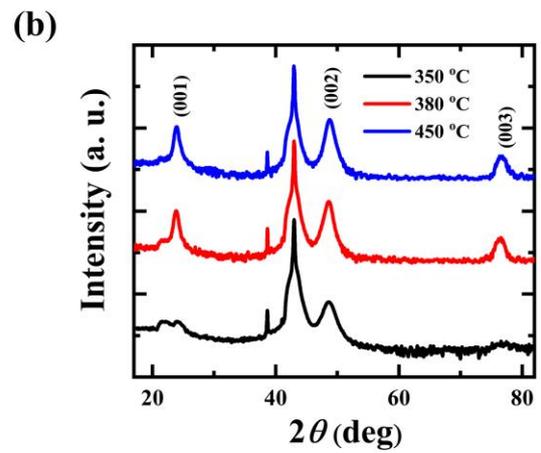

(b)

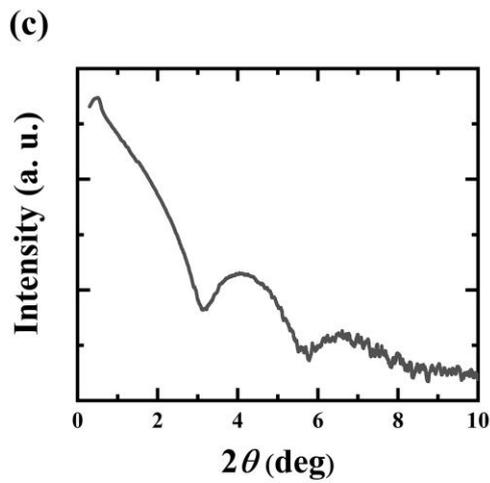

(c)

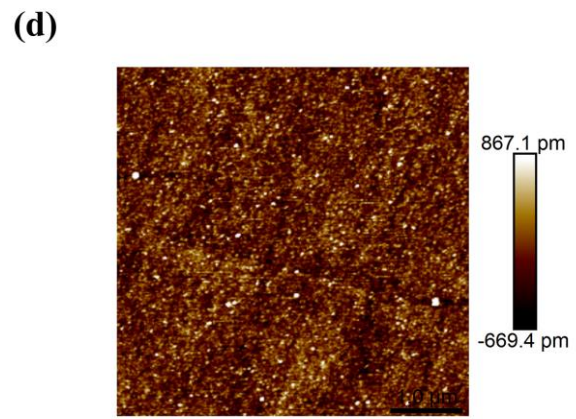

(d)

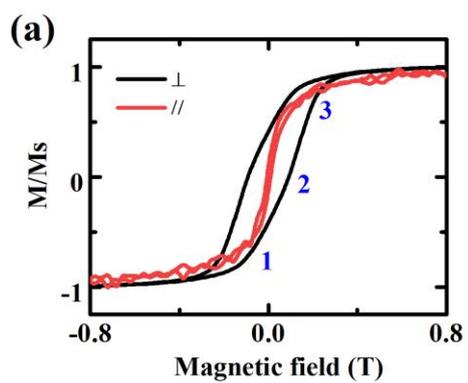
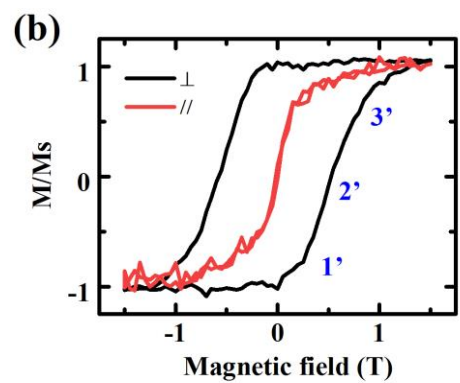
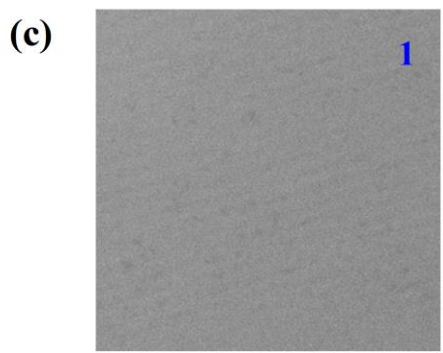
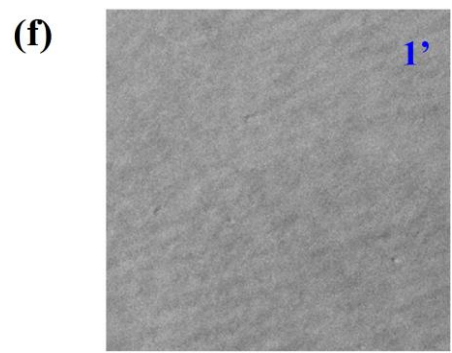
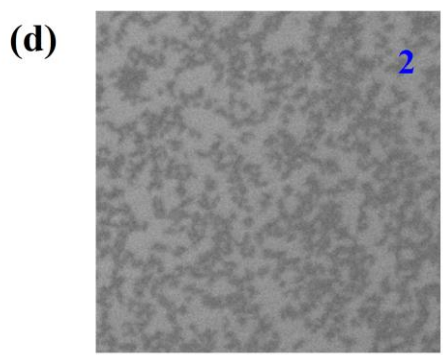
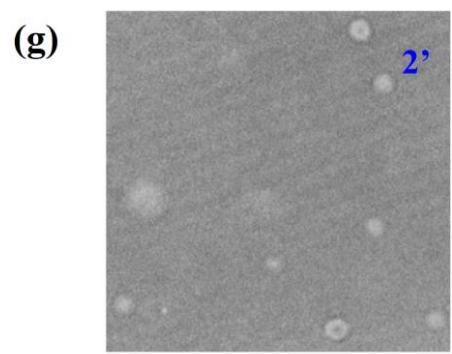
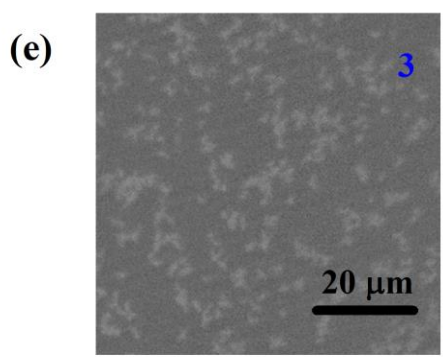
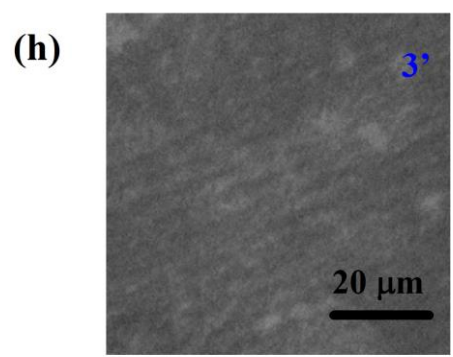

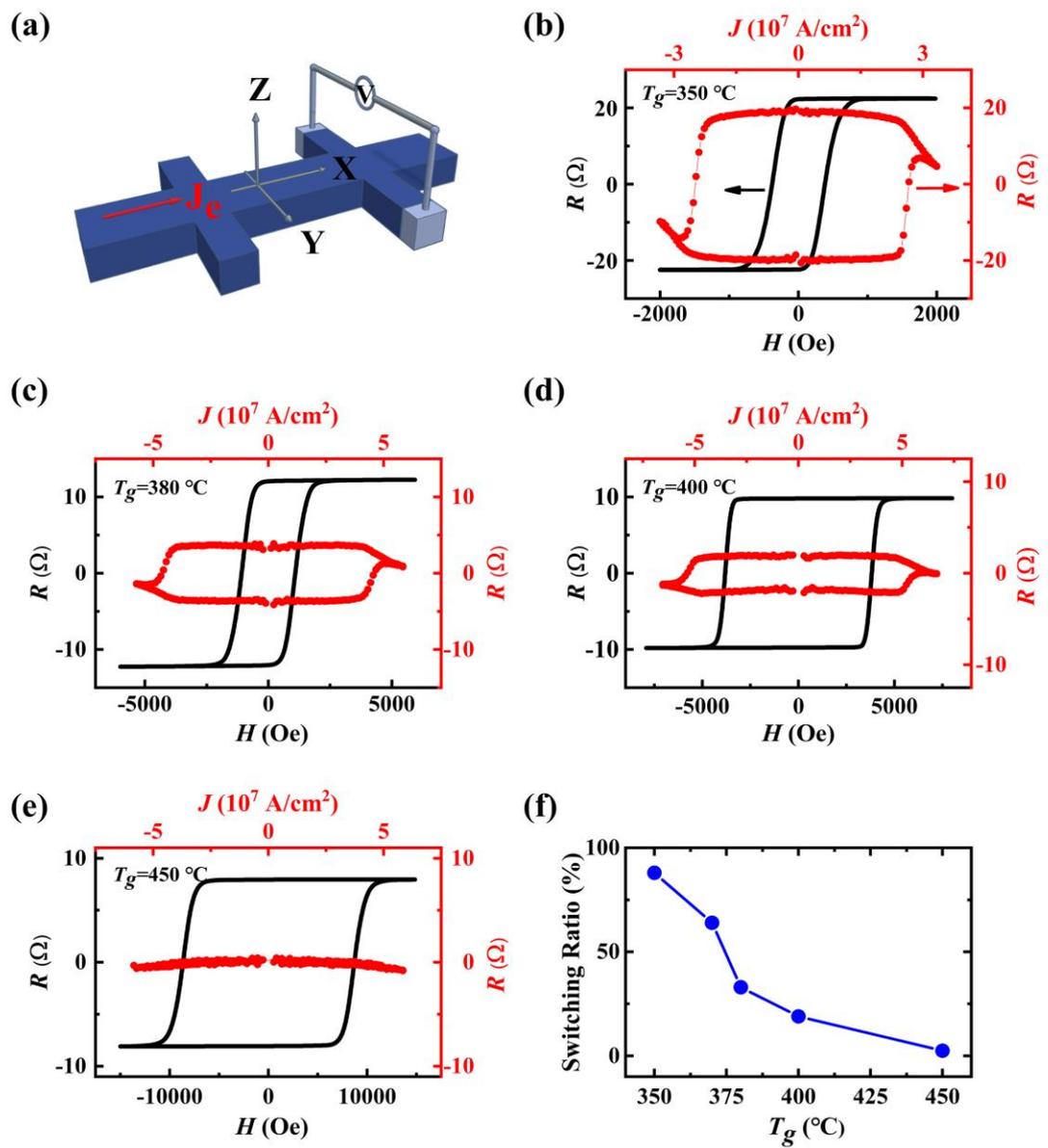

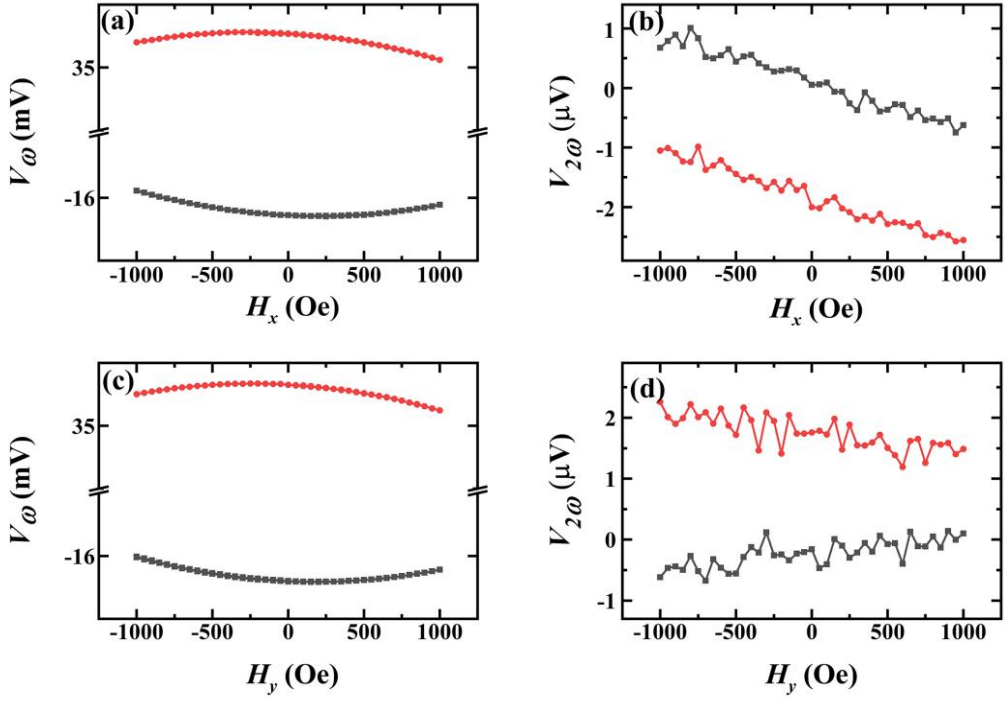
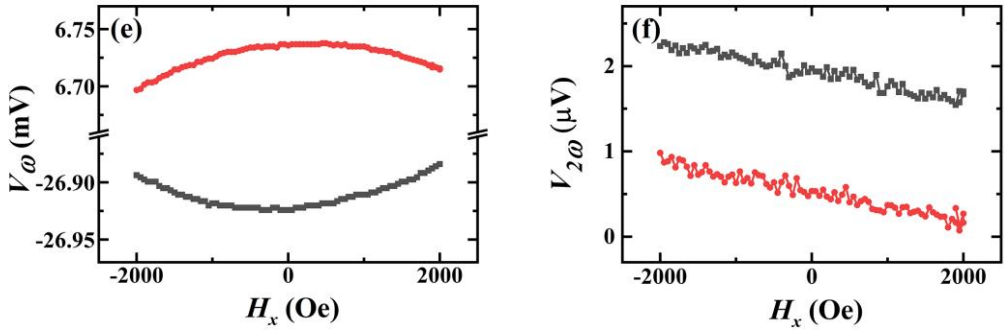

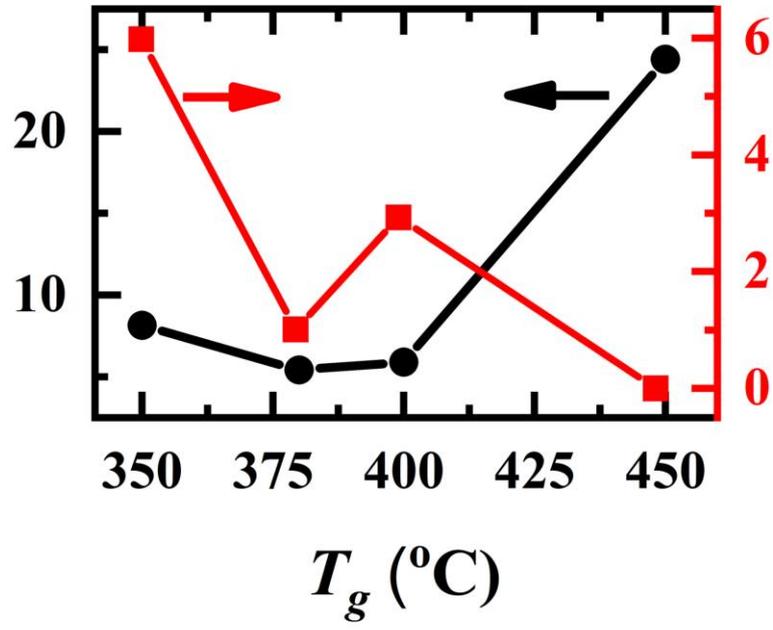

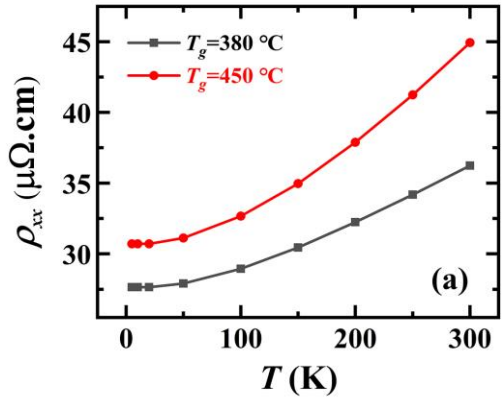 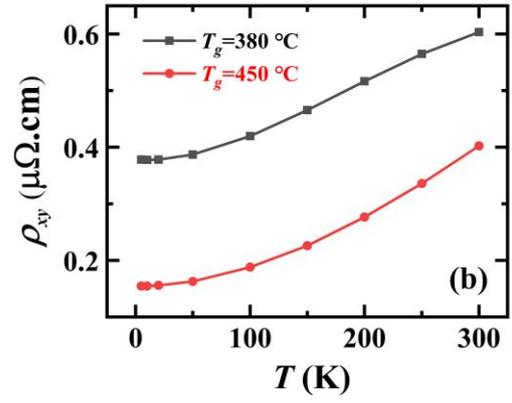
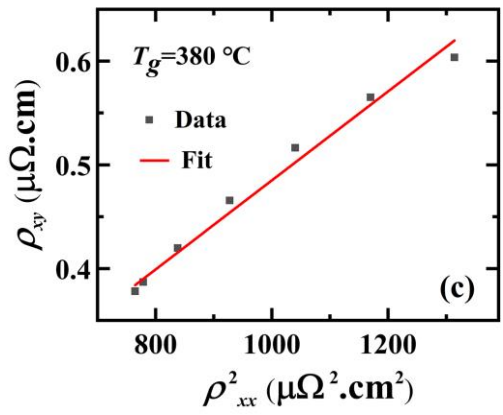 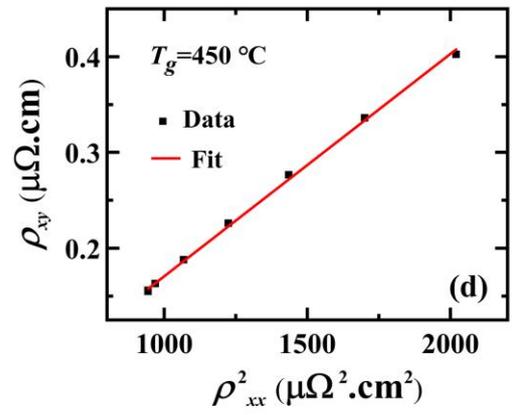

## Disorder dependent spin-orbit torques in $L1_0$ FePt single layer

S. Q. Zheng, K. K. Meng[*], Q. B. Liu, J. K. Chen, J. Miao, X. G. Xu, Y. Jiang[*]

*Beijing Advanced Innovation Center for Materials Genome Engineering, School of Materials Science and Engineering, University of Science and Technology Beijing, Beijing 100083, China*

We have also measured XRD of 3-nm-thick FePt films with $T_g$ =380 ℃ and $T_g$ =450 ℃ as shown in Fig. S1 to further investigate the structure properties and chemical ordering parameters of thinner films. The diffraction peak of FePt films could not be as strong as 20-nm-thick films in the main text. Besides the MgO peaks, one can observe the face-centered-tetragonal (fct) (001) and (002) peaks of FePt films in the spectrum, indicating the $L1_0$ ordering. The chemical ordering $S$ was calculated from the intensities of (001) and (002) peaks, and the values for films with $T_g$ =380 ℃ and $T_g$ =450 ℃ are 0.69 and 0.75, respectively.

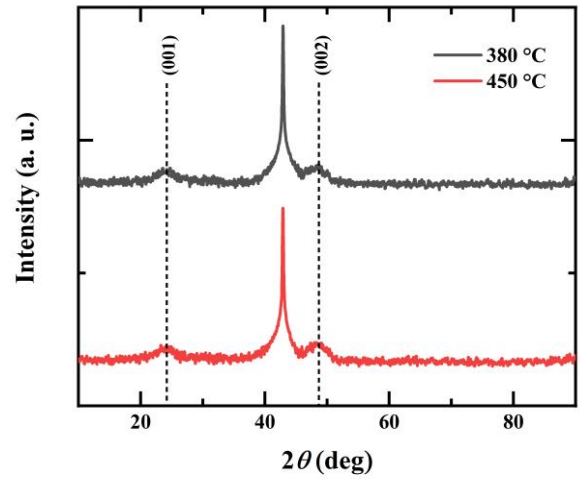

**Fig. S1.** X-ray diffraction of 3-nm-thick FePt films with different growth temperatures.

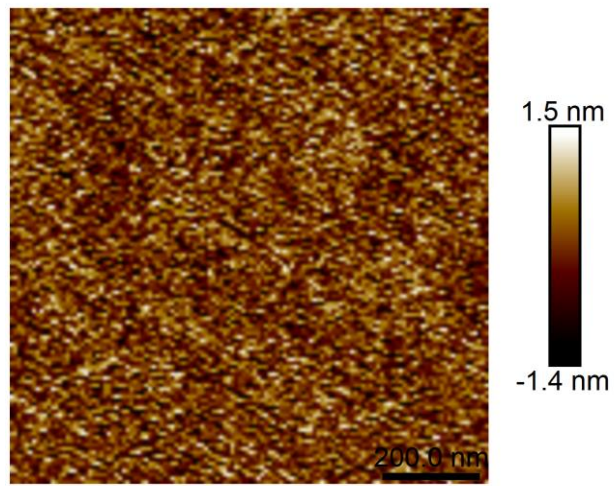

**Fig. S2.** Atomic force microscope of 3-nm-thick FePt films grown at 380 °C.

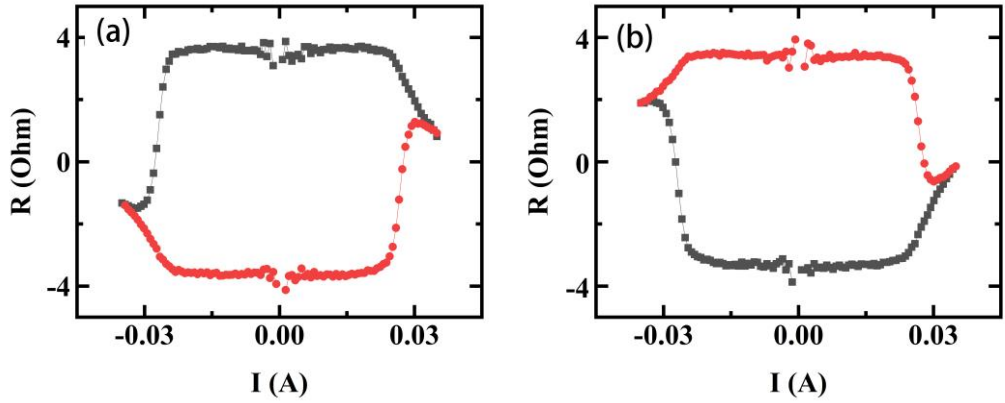

**Fig. S3.** The current induced magnetization switching behaviors of $L1_0$ FePt films with $T_g$=380 °C with $H_x$ =500 Oe (a) and $H_x$ =-500 Oe (b).

We have also measured the MOKE microscopy to characterize the magnetization switching driven by current for the films grown at 380 °C and 450 °C as shown in Fig. S4. It should be noted that, due to the MOKE system limitation, we can only carry out the experiments with relatively small current range. As the bright and dark contrasts in the Kerr images scale with the up and down magnetization components, respectively, the consecutive changes in Kerr contrast shown in the films grown at 380 °C confirm perpendicular magnetization switching in the FePt device. Moreover, the Kerr images for states at 14 mA and -14 mA (corresponding to $J=\pm2.3\times10^7$ A/cm$^2$) both show the formation of magnetic domains during the reversal process. The application of current triggers domain walls nucleation and expansion, leading to partially reversible magnetization switching. On the contrary, we have not found domain walls nucleation and expansion for the device with $T_g$ =450 °C in the same current range, revealing a relatively hard magnetization switching. Furthermore, the thermal effect is weak since the domain structures have not been varied with large current for the films grown at 450 °C.

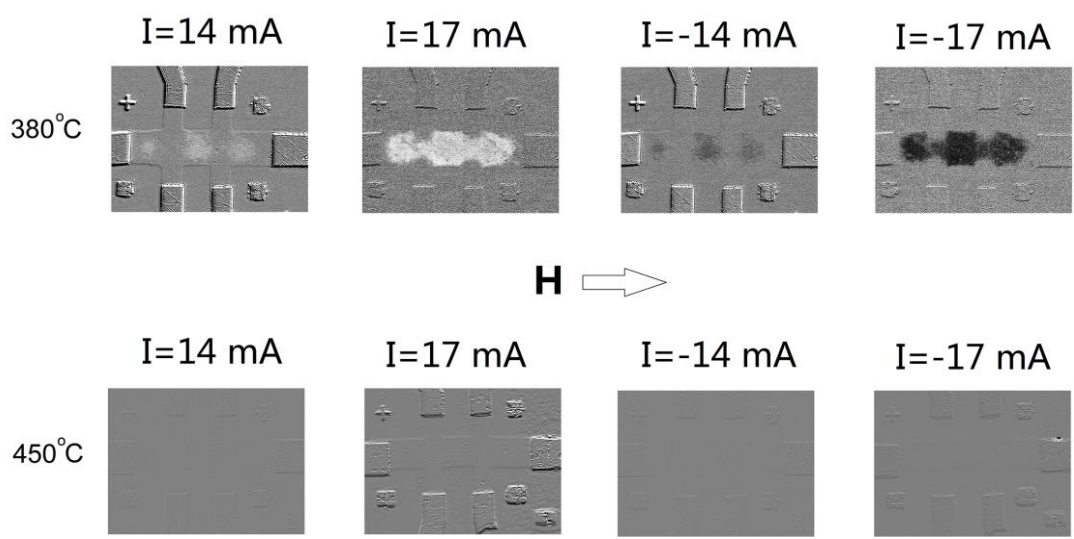

**Fig. S4**. Kerr images with different applied pulsed current for the two films.

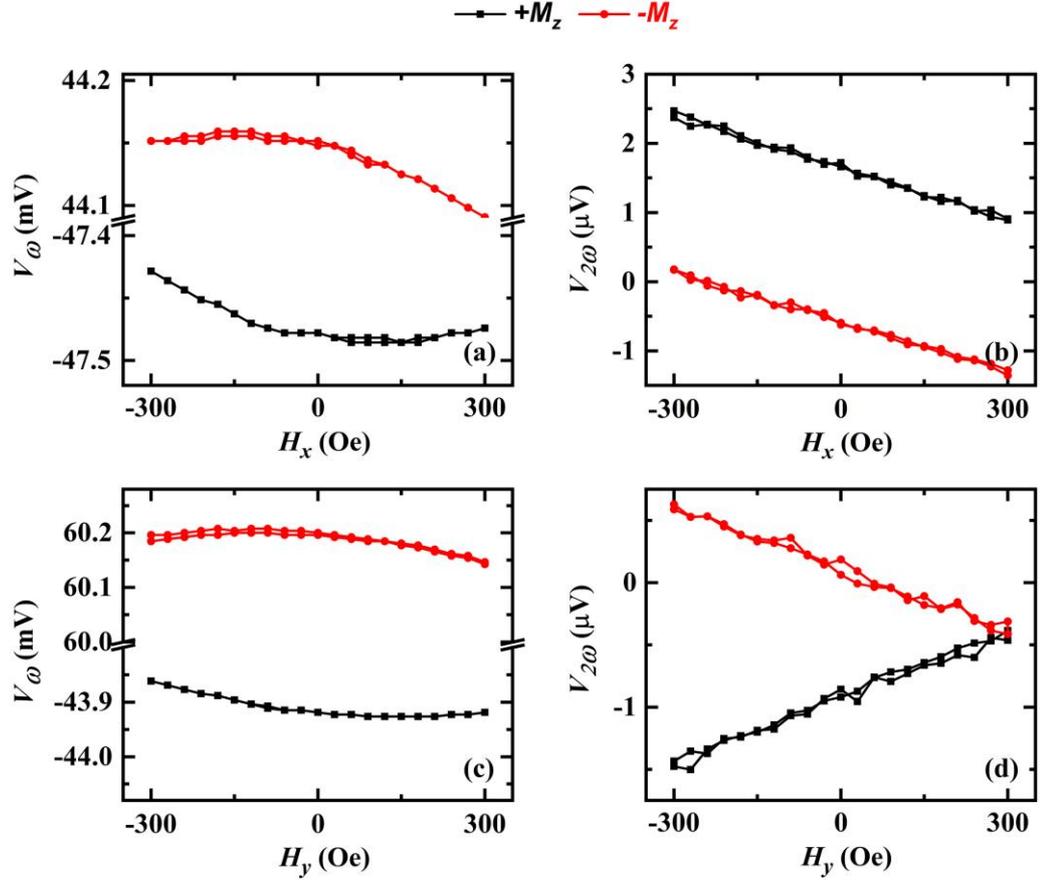

**Fig. S5.** The first $V_\omega$ and second $V_{2\omega}$ harmonic Hall voltages plotted against $H_x$ and $H_y$ in FePt film with $T_g$=350 °C.

The harmonic Hall measurement with different AC current have been applied for FePt films with $T_g$ =380 ℃ and $T_g$ =450 ℃. Considering the $H_F$ is negligible with increasing $T_g$, we have only plotted $H_D$-$I$ curves as shown in Fig. S6. The effective fields vary linearly with the current, indicating that the effects of Joule heating are negligible in the measured current range.

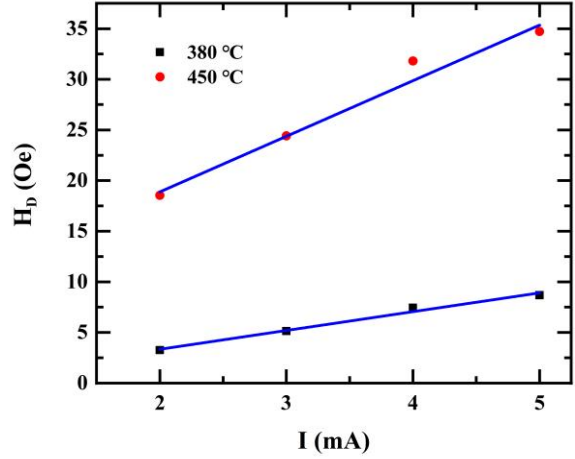

**Fig. S6.** Current-induced effective field $H_D$ versus applied AC current for FePt films with $T_g$ =380 ℃ and $T_g$ =450 ℃.

On the other hand, when the current flows through FePt films, the temperature gradient from different directions ($T_x$, $T_y$ and $T_z$) could generated during the measurements. The temperature gradient generates charge imbalances due to anomalous Nernst effect (ANE) and spin Seebeck effect (SSE), which could influence on the second harmonic Hall voltage during the measurements.

For the temperature gradients generated in the film plane ($T_x$ and $T_y$), their contributions to the second harmonic signals can be measured by sweeping magnetic field along the z-axis (normal to the film plane).[1] As shown in Fig. S7(a), we have measured $V_{2\omega}$ signals of FePt with $T_g$=450 ℃ by sweeping magnetic field along the z-axis. We can estimate the in-plane ANE contribution ($V_{ANE-IP}$) is 0.7 μV. The central value of the loop mainly comes from misalignment of the Hall voltage leads and the SSE,[1] which is the offset signal $V_{2\omega, Offset}$ (1.2 μV). As shown in Fig. S7(b), both $V_{ANE-IP}$ and $V_{2\omega, Offset}$ contribute to the second harmonic signals of FePt sample. On the other hand, $V_{ANE-IP}$ and $V_{2\omega, Offset}$ only give the second harmonic signal an overall constant offsets, which does not affect the slope of $V_{2\omega}$ in Fig. S7(b). Thus, the in-plane temperature gradients have no dramatic influence on the result of effective spin-torque fields.

In order to estimate the ANE contribution from $T_z$, we have measured the harmonic Hall voltages of FePt with $T_g$=450 ℃ with different magnetic fields rotating in *xy* plane based on the following equation in:[2]

$$R_{xy}^{2\omega} = [(R_{AHE}\frac{H_{AD}}{H_{ext}} + I_0\alpha\nabla T)\cos\phi + 2R_{PHE}(2\cos^3\phi - \cos\phi)\frac{H_{FL}+H_{Oe}}{H_{ext}}]$$

where both $R_{AD}^{2\omega}$ and $R_{\nabla T}^{2\omega}$ are proportional to cos$\varphi$, whereas $R_{FL}^{2\omega}$ is proportional to (2cos$^3\varphi$-cos$\varphi$). Therefore, by measuring the in-plane angle $\varphi$ (from x axis to y axis) dependence of $R_{xy}^{2\omega}$ at different external fields, the contribution of field-like torque (FL-SOT), damping-like torque (DL-SOT) and thermoelectric effects can be separated. Fig. S7(c) shows the angle $\varphi$ dependence of the second harmonic transverse resistances with the external field of 9 T. Considering that the contribution of FL-SOT ((2cos$^3\varphi$-cos$\varphi$)) are negligible, we only focus on the cos$\varphi$-like component. In order to further separate the thermal and DL-SOT contributions in the cos$\varphi$-like component of $R_{xy}^{2\omega}$, we verify the $R_{AD}^{2\omega}$+$R_{\nabla T}^{2\omega}$ is a linear function of 1/($H_{ext}$+$H_{dem}$-$H_{ani}$) as shown in Fig. S7(d), in which $H_{ext}$ is the external magnetic field, $H_{dem}$ and $H_{ani}$ is the demagnetizing field and perpendicular anisotropy field for FePt films. The slope of the fitted line shown in Fig. S7(d) can roughly represent the DL-SOT. The intercept could be associated with thermoelectric effects, which is close to zero. Thus, the ANE contribution from $T_z$ in our study can be neglected as well.

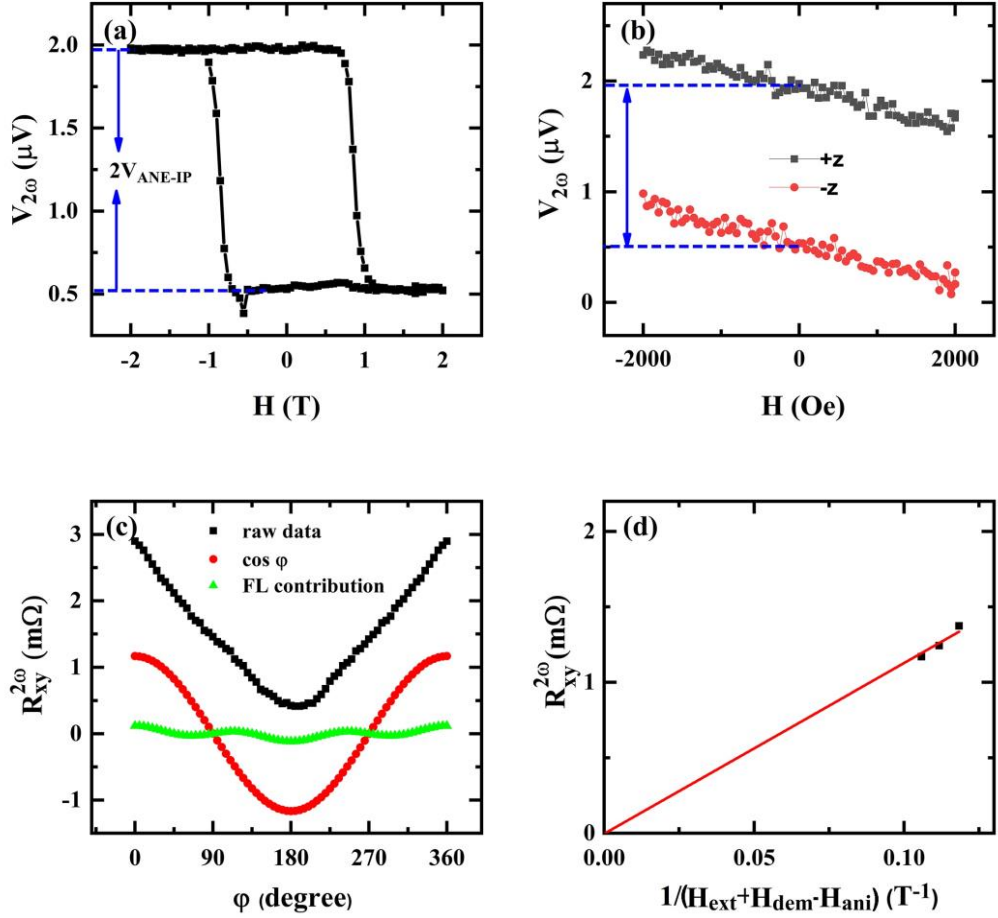

**Fig. S7.** (a) $V_{2\omega}$ plotted against large external fields $H_z$ in FePt with $T_g$=450 ℃ (b) $V_{2\omega}$ plotted against small external fields $H_x$ in FePt with $T_g$=450 ℃ (c) The in-plane angle (from x axis to y axis) dependence of the second harmonic transverse resistance in FePt with $T_g$=450 ℃ with the external field of 9 T. (d) $R_{AD}^{2\omega}+R_{\nabla T}^{2\omega}$ as a function of the inverse of the static fields acting against the current-induced field.

Another fitting approach for the AHE of FePt films is $\rho_{xy}=a'\rho_{xx0}+a''\rho_{xx}(T)+b\rho_{xx}^2$, where $\rho_{SK} = a'\rho_{xx0}+a''\rho_{xx}(T)$ represents the combined contribution from residual and phonon to skewed scattering [3]. We have also used this formula to fit the data as shown in Fig. S8. The first term $a'$ reveals the contribution from residual skew scattering, the second term $a''$ is the contribution from phonon scattering and the last term $b$ is the combined contribution from extrinsic side jump and intrinsic Berry curvature. Table SI shows the value of $a'$, $a''$ and $b$ of $L1_0$ FePt films with $T_g$=380 °C and $T_g$=450 °C. It can be found that the change of $b$ is small with varying $T_g$, while both residual skew scattering coefficient $a'$ and phonon scattering coefficient $a''$ change obviously, which have the same conclusion with the main text that the skew scattering can be dramatically varied due to the distribution, the type, and the density of disorder.

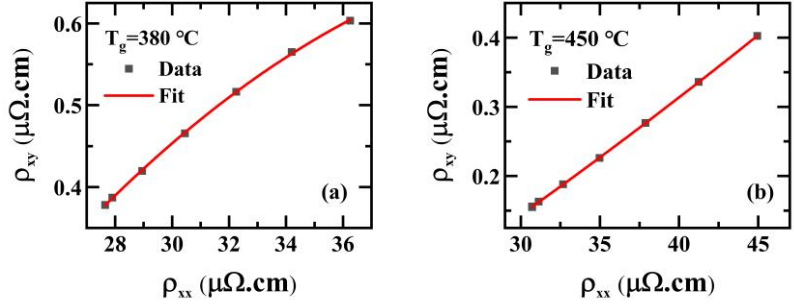

**Fig. S8.** The $\rho_{xy}$ as the function of $\rho_{xx}$ for $L1_0$ FePt film with $T_g$=380 °C (a) and $T_g$=450 °C (b). The red lines are fitted results using $\rho_{xy} = a'\rho_{xx0} + a''\rho_{xx}(T) + b\rho_{xx}^2$.

**TABLE SI.** Residual skew scattering coefficient $a'$, phonon scattering coefficient $a''$ and intrinsic coefficient $b$ for two $L1_0$ FePt films.

| $T_g$ (°C) | $a'$ | $a''$ | $b$ ($\mu\Omega^{-1}cm^{-1}$) |
|---|---|---|---|
| 380 | -0.04655 | 0.08604 | 9.33949E-4 |
| 450 | -0.00914 | 0.01205 | 7.02803E-4 |